\documentclass[showpacs,amsmath,amssymb,twocolumn,prl, floatfix]{revtex4}
\usepackage[dvips]{graphicx}
\input{epsf}

\begin{document}

\title{Regular modes in rotating stars}

\author
{Micka\"el Pasek$^{1,2,3,4}$}
\author
{Bertrand Georgeot$^{2,3}$}
\author
{Fran\c{c}ois Ligni\`eres$^{1,4}$}
\author
{Daniel R. Reese$^5$}
\affiliation{$^1$CNRS; IRAP; 14, avenue Edouard Belin, F-31400 Toulouse, France\\
$^2$ Universit\'e de Toulouse; UPS; Laboratoire de
 Physique Th\'eorique (IRSAMC); F-31062 Toulouse, France\\
$^3$ CNRS; LPT (IRSAMC); F-31062 Toulouse, France\\
$^4$ Universit\'e de Toulouse; UPS-OMP; IRAP; Toulouse, France\\
$^5$LESIA, CNRS, Universit\'e Pierre et Marie Curie, Universit\'e Denis Diderot, Observatoire de Paris, 92195 Meudon Cedex, France}

\date{May 27, 2011}








\begin{abstract}
Despite more and more observational data, stellar acoustic oscillation
modes are not well understood as soon as rotation cannot
be treated perturbatively. In a way similar to semiclassical theory in quantum physics, we use acoustic ray
dynamics to build an asymptotic theory for the subset of regular
modes
which are the easiest to observe and identify. Comparisons with 2D numerical
simulations of oscillations in polytropic stars show that both the frequency and amplitude distributions of these modes
 can accurately be described by an asymptotic theory for almost all rotation rates.
  The spectra are mainly characterized 
by two quantum numbers; their extraction from observed spectra should enable one to obtain information about stellar interiors.
\end{abstract}
\maketitle

Stars being far-away objects, the types of
information that can be obtained from them are necessarily limited. One of the most important
corresponds to luminosity variations, which can reflect the passing
of a planet or intrinsic modulations in the light emitted by the star.
In particular, the domain of asteroseismology studies stellar oscillation
modes, which create periodic variations of
the luminosity which can be detected \cite{Saio}.
For the Sun, these modes have been theoretically
constructed and successfully compared with observations, leading
to detailed information on the Sun's internal structure.
However, this theory requires the star to be nearly spherically symmetric,
an assumption clearly violated for rapidly rotating stars \cite{Monnier}. With the launch
of the recent space missions COROT and Kepler \cite{Saio}, 
oscillation spectra of
rapidly rotating stars are observed with great accuracy.  This concerns mainly the stars more massive than the Sun that belong to the main
sequence of the Hertzsprung-Russel diagram. In order to access their 
internal
structure through a seismic diagnostic, it is thus essential to 
understand the
oscillation spectra of rapidly rotating stars

\begin{figure}
\includegraphics[width=0.95\linewidth]{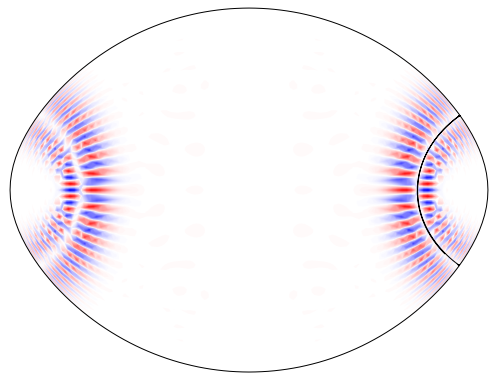}

\caption{(Color online) Pressure amplitude $P\sqrt{d/\rho_0}$ on a meridian plane for a polytropic model of stars, with $d$ the distance to the rotation axis and $\rho_0$ the equilibrium density. The mode shown corresponds to $n=46, \ell=1$ and $m=0$ at a
rotation rate of $\frac{\Omega}{\Omega _K} = 0.783$, with
$\Omega_K =(GM/R_{eq}^3)^{1/2}$ being
the limiting rotation rate for which the centrifugal acceleration equals the gravity at the equator,  $M$ the stellar mass and $R_{eq}$ the equatorial
radius. Colors/grayness denote pressure amplitude, from red/gray (maximum positive value) to blue/black (minimum negative value) through white (zero value). The thick black line on the right is the central periodic orbit $\gamma$ of the island.
}
\label{fig1}
\end{figure}

Accurate computations of acoustic modes fully taking into account the effects of rotation on stellar oscillations
have only recently been performed for rotating stars \cite{Lign} (an example is shown in Fig.~\ref{fig1}).
Such stationary patterns
of acoustic waves can be described asymptotically through
their short-wavelength limit, in the same way
as classical trajectories can describe quantum or electromagnetic waves
in this limit \cite{Goug}.  These acoustic rays obey
Hamiltonian equations of motion. In \cite{LG}, their dynamics was investigated for a polytropic stellar model, showing that the tools from the fields of classical and quantum chaos enable us to understand the behavior of modes in rapidly rotating stars. Indeed, for increasing rotation rates, the dynamics undergoes a transition from an integrable to a mixed system,
where chaotic and stable zones coexist in phase space.  The asymptotic theory built for slowly rotating stars, which does not take these effects into account, cannot thus be applied at
high rotation rates. In the latter regime, it was shown that
the spectrum of acoustic oscillations
can be divided into several subspectra corresponding to regular and chaotic
zones in phase space in a way similar to what happens in quantum
chaos systems \cite{Berr}.

As already demonstrated for the Sun and solar-type stars, a quantitative asymptotic theory
is crucial to extract information from the observed spectra as it links
the behavior of oscillation
modes to physical properties of the star \cite{Chri}.   Furthermore, the complexity of the observed spectra usually requires prior knowledge of an asymptotic theory in order to correctly identify the frequency peaks in the data with specific oscillation modes.
In this paper, we present for the first time such a quantitative theory at almost all rotation rates for
a specific subset of modes, which should be among the easiest to obtain
from observations. Indeed, we focus on a series of modes centered around the
largest stable island of the system, and systematically build them using the parabolic equation method \cite{babich}.
This method was successfully applied to light in dielectric
cavities \cite{Stone}, electronic resonators in a magnetic field \cite{Zalipaev} and quantum chaos systems \cite{Schomerus}.
The results of the asymptotic modes are then compared with numerical computation of oscillations in a polytropic star,
showing that the theory correctly describes the
modes even in the bounded frequency range of stellar oscillations.

\begin{figure}
\includegraphics[width=0.8\linewidth]{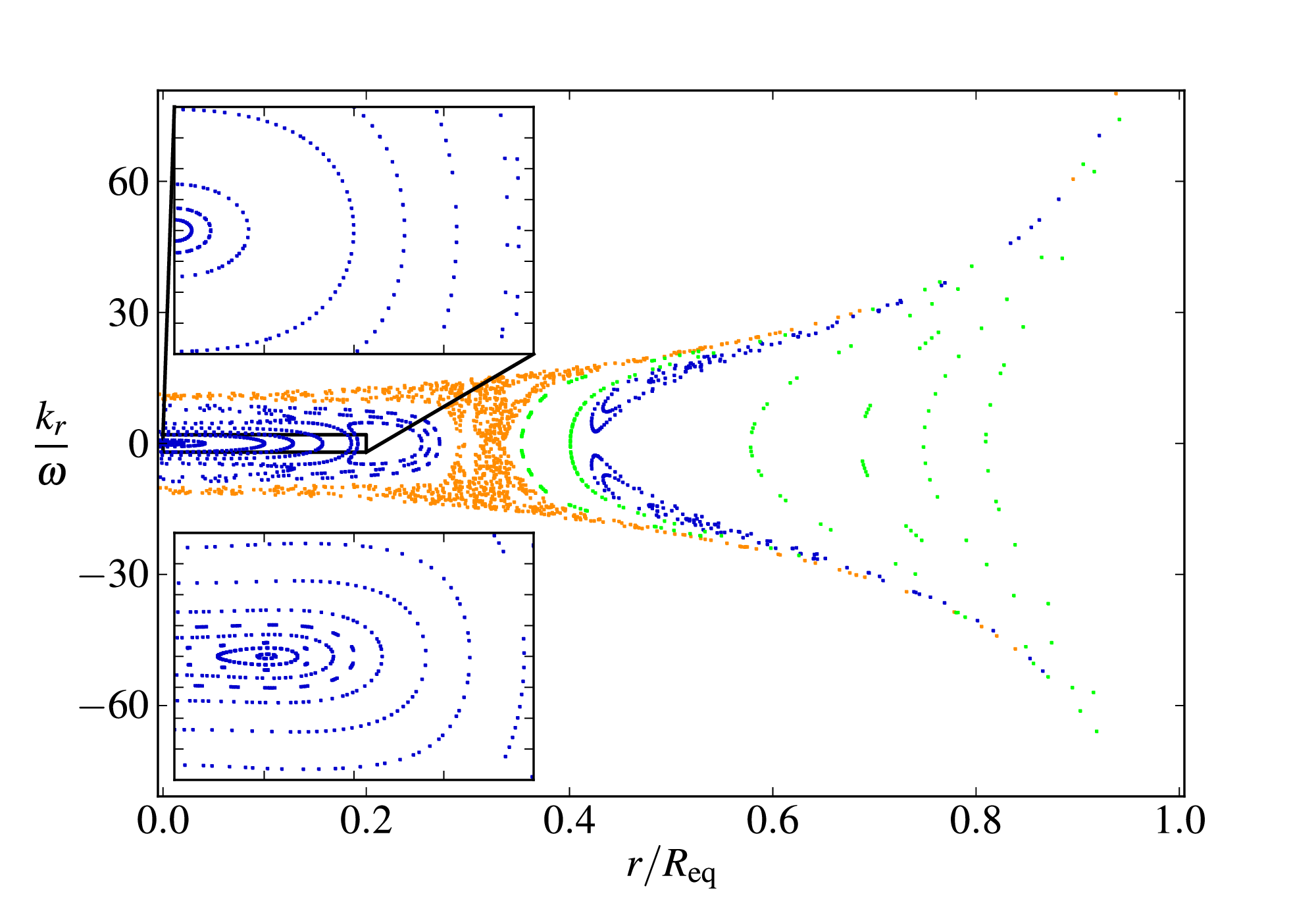}\\
\includegraphics[width=0.8\linewidth]{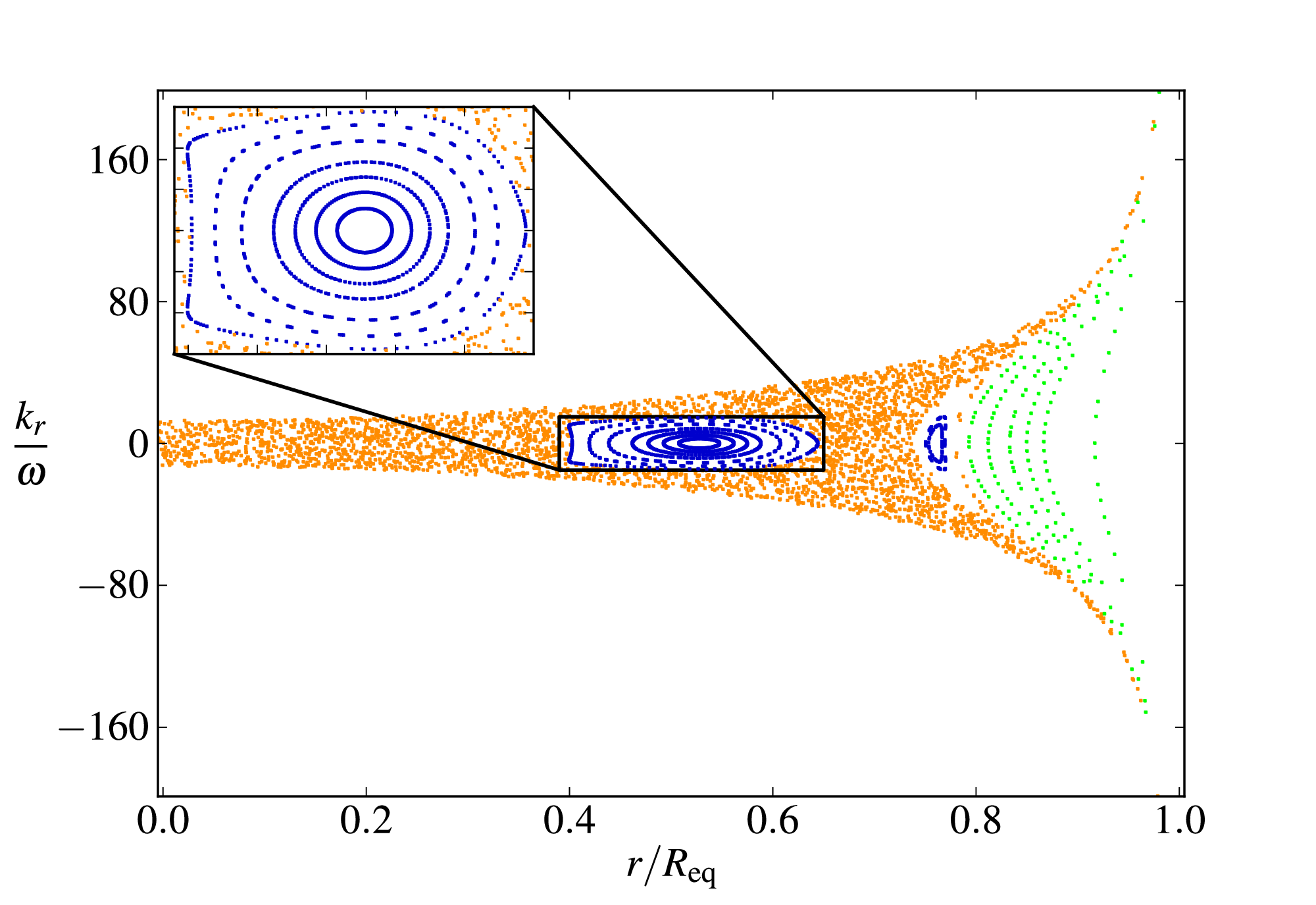}

\caption{(Color online) Surfaces of section at rotations $\frac{\Omega}{\Omega _K} =0.224$ (top) and
$\frac{\Omega}{\Omega _K} = 0.589$ (bottom) for $m=0$. Each dot represents the crossing of an acoustic ray with the equatorial half-plane, $r$ being the radial coordinate and $k_r$ the associated momentum. Orange/light gray denote a chaotic ray, green/dark gray a whispering gallery ray, blue/black a stable island ray (see text). Upper insets are close-ups of the main stable island. Lower
inset in the top figure shows a close-up of the main island at 
$\frac{\Omega}{\Omega _K} = 0.262$, just after the bifurcation.
}
\label{fig2}
\end{figure}

The study of ray dynamics in \cite{LG} showed that for a wide range of rotation rates, three main types of phase space zones with different dynamics can be defined (see Fig.\ref{fig2}):
1) regular 
structures built around stable periodic orbits (stable islands); 2) whispering gallery rays close to the surface; 3) chaotic zones with ergodic rays in the remaining parts of phase space. Figure \ref{fig2} shows that for $m=0$ (axisymmetric modes) the main island undergoes
a bifurcation from one island centered on the rotation axis to two islands which move away from the rotation axis
as the rotation rate increases. Each phase space region gives rise to a well-defined subspectrum of modes localized inside the region. The whispering gallery modes are essentially unobservable in real stars since the disk-average
leads to a very small contribution in observed spectra \cite{LG}.  Chaotic modes can have visible contributions, but the associated spectra, although they
can display well-defined statistical properties, cannot be described by a few quantum
numbers. In contrast, the stable island modes give rise to very regular sequences of frequencies described by a few parameters which can be potentially extracted from observed spectra. 
These modes (an example is shown on Fig.~1) can be characterized on a meridian plane by the number $n$ of nodes
along the central periodic orbit $\gamma$ and the number $\ell$ of nodes transverse to $\gamma$.

To describe asymptotically these island modes, we start from the equation of
acoustic waves in stars.  We neglect the Coriolis force, which is known to
be negligible in the high-frequency regime since the Coriolis force time scale ($1/(2 \Omega)$) is much longer than the mode period \cite{Lign}. We also
neglect the perturbations of the gravitational potential, since they are produced by the density fluctuations and tend to cancel out as the number of nodes of the density distribution increases for high-frequency modes, as has been numerically checked for non-rotating stars \cite{Chri2}. Finally, we use the adiabatic approximation which is known to be a very good approximation to compute frequency modes in non-rotating stars. Indeed, it is accurate enough to interpret the solar acoustic modes
despite the fact that
these frequencies are determined to high accuracy \cite{Chri}.
In the linear approximation, this gives rise
to a Helmholtz-type equation.  Using the cylindrical symmetry of the system with respect to the rotation axis,
one can rewrite this equation as a two-dimensional problem
\begin{equation}
 -c_s ^2 \Delta \Phi_m + \left[ \omega_c ^2 + \frac{ c_s ^2 (m^2 - \frac{1}{4})}{d^2} \right] \Phi_m = \omega ^2 \Phi_m .
\label{helm}
\end{equation}
Here $\Phi_m$ is the mode amplitude scaled by the square root of the distance $d$ to the rotation axis, $c_s$ is the sound velocity (which depends on the
location inside the star), $\omega$ is the frequency of the mode,
 and ${\omega}_c$
is the cut-off frequency whose sharp increase in the outermost layers of the star leads to the
 reflection of acoustic waves and thus the formation of modes in a bounded frequency range $\omega \le (\omega_c)_{\rm max}$. The integer
$m$ is the quantum number corresponding
to the quantization of angular momentum along the rotation axis.
To construct the island modes centered on the stable periodic orbit $\gamma$ 
of length $L_{\gamma}$,
we rewrite Eq.~(\ref{helm}) in the coordinates $(s,\xi)$ centered on $\gamma$, 
with $s$ the coordinate along $\gamma$ and $\xi$ the transverse coordinate. The
parabolic equation method \cite{babich} assumes that the solutions have a longitudinal scaling in $1/\omega$ and a transverse
scaling in $1/\sqrt{\omega}$.
We  introduce the WKB ansatz:
\begin{equation}
 \Phi_m (s,\xi) = \exp(i \omega \tau ) U_m (s, \xi,\omega),
\end{equation}
with $d\tau=ds/\tilde{c}_s$, using the renormalized sound velocity defined by
$\tilde{c}_s^2=c_s^2 \omega ^2/(\omega ^2-\omega_c ^2 - \frac{ c_s ^2 (m^2 - \frac{1}{4})}{d^2})$.

An expansion in powers of $\omega$ yields a series of equations; keeping
terms of order $\omega$ and introducing the variable $\nu = \sqrt{\omega} \xi$ and the function $V_m = U_m / \sqrt{\tilde{c}_s(s)} $ give the parabolic equation:

\begin{equation}
\frac{\partial ^2 V_m}{\partial \nu ^2} + 2 i \frac{1}{\tilde{c}_s (s)} \frac{\partial V_m}{\partial s} - K(s) \nu ^2 V_m = 0,
\end{equation}
where 
$K(s) = \frac{1}{\tilde{c}_s(s)^3} \frac{\partial ^2 \tilde{c}_s}{\partial \xi ^2}\Big |_{\xi=0}$.
The equation in the transverse coordinate $\nu$ is similar to the harmonic oscillator in 
quantum mechanics, with an additional term depending on the longitudinal coordinate.
The ground state is of the form:
$ V_{m}^0  = A(s) \exp \left[ i \frac{\Gamma(s)}{2} \nu ^2 \right]$
and obeys the two equations:
$ \frac{1}{\tilde{c}_s(s)}\frac{d \Gamma(s)}{d s} + \Gamma(s) ^2 + K(s) = 0$ and $\frac{1}{A(s)} \frac{d A (s)}{d s} = - \frac{\tilde{c}_s (s)}{2} \Gamma(s)$.
Defining $z(s)$ through
 $ \Gamma(s) = \frac{1}{z(s) \tilde{c}_s(s)}\frac{d z (s)}{d s}$ implies
that  $A (s) = 1/\sqrt{z(s)}$ and $z(s)$ should satisfy
the following system of equations:

\begin{equation}
\frac{1}{\tilde{c}_s(s)} \frac{d z}{d s}  = p ~,~ \frac{1}{\tilde{c}_s(s)} \frac{d p}{d s}  =  - K(s) z~.
\label{p}
\end{equation}

Equations (\ref{p}) are periodic in the variable $s$ with period $L_{\gamma}$,
and according to Floquet theory there exists an operator describing the
evolution over one period.
To construct it, one uses the fact that
 system (\ref{p}) corresponds to the Hamilton equations associated
with the Hamiltonian
$H = \frac{p ^2}{2} + \frac{K(s) z^2}{2}$.  It can be shown that the same equations (with $z$ and $p$ real) describe the acoustic ray in the vicinity
of the central periodic orbit (via a normal form approximation), $z$ being the transverse deviation from $\gamma$  and $p$ the associated momentum.   
We linearize the motion around the periodic
orbit and construct the monodromy matrix which describes this linearized
motion from one point to its image after one period:
 \begin{equation}
\left[ \begin{array}{c}
z(s+L_\gamma) \\
p(s+L_\gamma)
\end{array} \right]  = M 
\left[ \begin{array}{c}
z(s) \\
p(s)
\end{array} \right] ~.
\end{equation}
For the mode to be univalued, $ V_{m}^0$ should be the same after one period up to a global phase and thus $z$ and $p$ should correspond to an eigenvector of $M$.
As $\gamma$ is stable, the matrix $M$ is conjugate to a rotation matrix
and has two eigenvalues $ e ^{\pm i \alpha}$, with $\alpha$ in $[0,2\pi[$. The
corresponding eigenvectors are complex conjugate, 
only one of them giving
the physical solution exponentially decreasing at large $\nu$.

The modes of higher frequency can be constructed as for
the harmonic oscillator from $ V_{m}^0$ using standard methods
from quantum mechanics; the result, up to a normalization constant, is equivalent to multiplying 
$ V_{m}^0(s,\nu)=z^{-\frac{1}{2}}\exp \left[ i \frac{\Gamma}{2} \nu ^2 \right]$
 by a function containing the Hermite 
polynomials of order $\ell$ noted $H_\ell$:
\begin{equation}
 V_{m}^\ell(s,\nu) = \left(\frac{\bar{z}}{z}\right)^{\ell/2} H_\ell (\sqrt{\mbox{Im}\Gamma} \nu) z^{-\frac{1}{2}} \exp \left[ i \frac{\Gamma}{2} \nu ^2 \right]~.
\label{hermite}
\end{equation} 

Again, for the mode to be univalued, the global phase accumulated after one period should be
a multiple of $2\pi$.
This phase is  $ \exp\left(- i \pi\right)$ $ \exp \left(- i (2\pi N_r+\alpha)/2 \right)$ $  \exp \left(- i (2\pi N_r+\alpha)\ell \right)$ $\exp\left( i \omega \oint_\gamma \frac{ds}{\tilde{c}_s} \right)$.
The first two phases
correspond to the so-called Maslov indices \cite{Goug,gutz} and count the number of caustics encountered in the longitudinal and transverse motions. The number $N_r$ keeps track of the number of times the trajectory solution of Eq.~(\ref{p}) makes a complete rotation around $\gamma$ in phase space, and can be evaluated from ray simulations. This implies:
\begin{equation}
  \omega_{n,\ell,m} = \frac{1}{\oint_\gamma \frac{ds}{\tilde{c}_s}} \left[2 \pi (n+\frac{1}{2}) + \left( \ell + \frac{1}{2} \right) (2\pi N_r+\alpha) \right]~.
\label{eqmaj}
 \end{equation}

 The regular subspectrum
is thus essentially described by two quantities, $\delta n =\frac{2\pi}{\oint_\gamma \frac{ds}{\tilde{c}_s}} $ and $\delta \ell =\frac{2\pi N_r+\alpha}{\oint_\gamma \frac{ds}{\tilde{c}_s}} $ (which depend on $m$).  This corresponds to the empirical formula found in \cite{Lign} for $m=0$ from numerical simulations. 
The quantities ${\delta} n$ and  $\delta \ell$ probe the sound velocity along the path of the periodic orbit and its transverse derivatives.  Indeed, an explicit expression of  $\alpha$ in terms of such transverse derivatives can be derived \cite{gutz}. Equation (\ref{eqmaj}) is valid asymptotically for $n$ large and $\ell \ll n$.
As observable modes in real stars cannot be too high in frequency, we have
checked numerically the validity of this formula for moderately high values of $n$.

\begin{figure}
\includegraphics[width=0.95\linewidth]{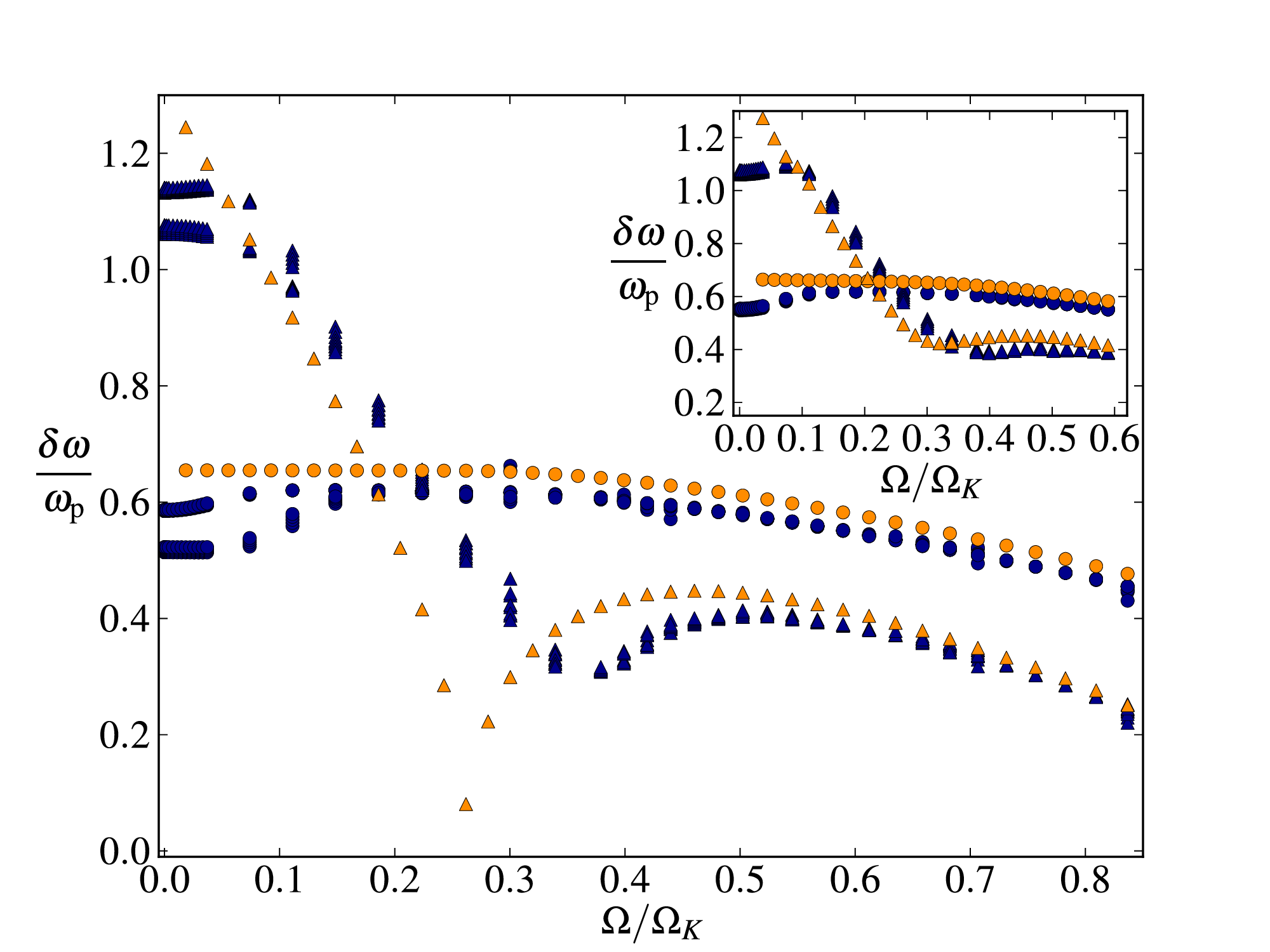}
\caption{(Color online) Comparison between actual regularities of regular modes
 and theoretical predictions for $m=0$ and different values of $\Omega / \Omega _K$ ($\omega_p=\sqrt{G M/R_p^3}$ with $R_p$ the polar radius).
Circles: $\delta n$, triangles: $\delta \ell$, orange/light gray: theory, blue/dark gray: numerical results.  
Numerical results correspond to different sets of values of
$\delta n$ and $\delta \ell$, with $n$ between $42$ and  $51$ and $\ell=0,1$.
As $m=0$, only modes symmetric with respect to the 
rotation axis should be retained, and thus the theoretical value of $\delta \ell$
is multiplied by two below the bifurcation point. Close to $\Omega =0$, the separation of numerical values in two groups corresponds to the small separation in
Tassoul's theory \cite{Chri}. 
Inset: Same for $m= 1$.
}\label{fig4}
\end{figure}

In Fig.\ref{fig4} we plot the numerically computed $\delta n$ and 
$\delta \ell$ values vs.~the theoretical ones 
for a large range of rotation rates.  We restrict ourselves
to the case $|m| \leq 1$ which is the most common in observational data.
Numerical modes
were obtained using a code that computes adiabatic modes of rotating polytropic stars  as in \cite{Lign},
and selecting the island modes through their phase space locations.  Theoretical values were obtained from
the theory explained above, estimating the monodromy matrix entries
by following classical trajectories in the vicinity of the periodic orbit,
using the fact previously noted that 
Eq.~(\ref{p}) describes the deviation of a nearby classical trajectory from the
central orbit $\gamma$.
We checked that the results were not sensitive to the choice of the
trajectory inside the island.  The results of Fig.\ref{fig4} 
show that a good agreement exists between numerical and theoretical
regularities, except close to $\Omega=0$ where Tassoul's asymptotic theory applies \cite{Chri}. For $\delta n$, the agreement is good over the whole
range of rotation. For $\delta \ell$, the agreement is good at large
and low rotation, but degrades in the range $[0.25,0.35]$ for $m=0$.  
We attribute
this discrepancy to the fact that, as seen in Fig.\ref{fig2}, 
the periodic orbit 
of the main stability island undergoes a bifurcation in this range,
from one stable central orbit to two stable orbits on each side and a central
unstable one.  It is known that in such a case, the normal form approximation
for the classical motion which is used in the parabolic equation method
should be modified by different uniform formulas \cite{bifurcation}.
Thus in the vicinity of the bifurcation the method is expected not to give accurate results.   This picture is confirmed by the inset
of Fig.\ref{fig4}, which shows that in the case $m= 1$, where there is no such bifurcation, agreement is good for $\delta \ell$ over the whole range of
$\Omega$ values. We note that other bifurcations are present in the system 
which create additional stable and unstable orbits in the vicinity of the central one, but they do not seem to affect the results for the relatively low-frequency modes we consider.
We note also that Eq.~(\ref{eqmaj}) predicts degeneracies at rational values of 
$\alpha/\pi$.  These degeneracies can be avoided crossings or true degeneracies if the modes belong to different symmetry classes. We have checked that it
actually enables us to predict such occurrences.  We note that while the theory neglects the Coriolis force and perturbations of the gravitational potential, the numerical modes were computed taking into account both effects. The good agreement seen in Fig.\ref{fig4} confirms that these processes can safely be neglected in this regime. We also remark that recent analysis of numerically computed modes in realistic, non-polytropic, differentially rotating stellar models show the emergence of formulas similar to Eq.~(\ref{eqmaj}) for specific subsets of modes \cite{reesebis}.

\begin{figure}

\includegraphics[width=0.95\linewidth]{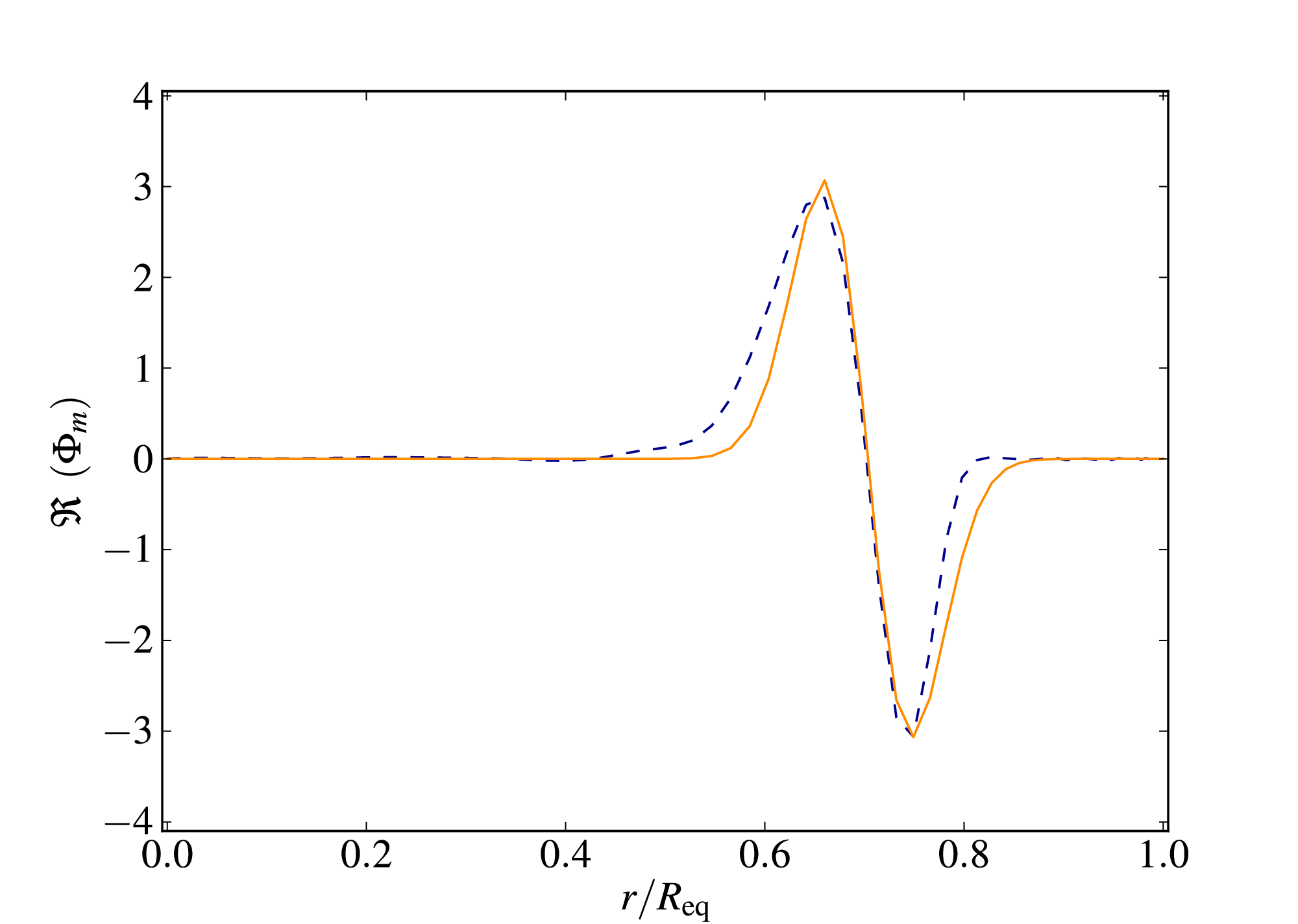}

\caption{(Color online) Amplitude distributions (real part of $\Phi_m$) on the equator
for a theoretical and a numerical mode. Blue dashed line: numerical mode (same as in Fig.\ref{fig1});
orange continuous line: theoretical mode, from Eq.~(\ref{hermite}).
}
\label{fig5}
\end{figure}

Not only does the parabolic equation method give the frequencies
of the modes, but it also yields their amplitude distribution.  Indeed,
the eigenvector of the monodromy matrix gives $\Gamma(s)$, which enables us
to construct an approximation of the mode itself using Eq.~(\ref{hermite}).
Comparisons between theoretical and numerical modes show that the modes
are well approximated by the theory (see an example in Fig.\ref{fig5}), although sometimes small oscillations due
to interference between different modes are not well reproduced by
the theory.

In conclusion, we have shown that the parabolic equation
method enables us to build an asymptotic theory for the most visible of the regular acoustic modes of
a star rotating at arbitrary rotation rates except for very slow rotation,
where Tassoul's theory \cite{Chri} already applies. Comparisons with numerical
computations of oscillations in a stellar model show that the asymptotic theory
gives a good description of the frequency differences and 
amplitude distributions, except for $m=0$ at a specific rotation rate where a 
bifurcation takes place and a more refined theory is needed.  The spacings
$\delta n$ and $\delta \ell$ which describe the frequency distribution of 
this type of modes can be expressed in terms of internal characteristics of
the star.
Our results should enable one to use data from recent 
space missions such as COROT and Kepler to extract information about the
observed stars and use this information to build more accurate stellar models. 

\begin{acknowledgments} We thank J. Ballot for his help at
various stages of this work, the ANR project SIROCO for funding and
CALMIP and CINES for the use of their supercomputers. DRR acknowledges 
support from the CNES through a post-doctoral fellowship.
\end{acknowledgments}

\end{document}